\documentclass[12pt]{article}
\title{Constructing Point Form Mass Operators from Vertex Interactions}
\author{W. H. Klink\\Department of Physics and Astronomy\\
 University of Iowa, Iowa City, Iowa, USA}
\begin{document}
\maketitle
\begin{abstract}
A relativistic few-body theory is formulated using point form quantum mechanics,
in which all of the interactions are in the four-momentum operator and Lorentz
transformations are kinematic.  The four-momentum operator is written as a
product of mass and four-velocity operators, where the mass operator is the sum
of free and interacting mass operators.  Interacting mass operators are 
constructed from vertices, products of local field operators, evaluated at the
space-time point zero.  Matrix elements of such mass operators, evaluated on
four-velocity eigenstates of a truncated Fock space, which is the space of the
few-body theory, are shown to behave like relativistic potentials.  
 Examples for a simple
vertex are given\footnote{P.A.C.S. 21.30.Fe, 24.10Jv, 25.10.+s, 25.80.Dj}.
\end{abstract}
\section{Introduction}
One of the goals of few-body nonrelativistic nuclear physics has been to
use phenomenological potentials determined by two-body systems (for
example the deuteron and nucleon-nucleon scattering) to predict the
behavior of three or more body systems.  Similarly, current operators
determined by one and two body systems are used to predict form factors
for three or more body systems.  In the energy domain where nonrelativistic 
quantum mechanics is
valid this program has been quite successful \cite{a}.

To carry out a similar program at higher energies requires introducing
some form of relativistic quantum mechanics, not only because of the higher energies involved, but also
because particle ( for example pi meson) production becomes
important. One of the ways to generate a Poincar\'{e} covariant theory is via 
the so-called
Bakamjian-Thomas procedure \cite{b}\cite{c}, where interactions for the few body system
are put into a mass operator, which takes the place of the Hamiltonian. 
While this procedure is indeed relativistic, it
generally violates cluster separability
properties, wherein if two
clusters of interacting particles are widely separated, there should be
no interactions between the separated clusters.  It has been shown that
cluster separability properties can be satisfied by introducing
appropriate unitary operators \cite{d}, but this procedure is unwieldy,
and in any event it is not entirely clear how to extend it to deal with
particle production.

Quantum field theory on the other hand provides a setting in which
particle production occurs naturally, as does cluster separability. 
However when the Fock space on which the quantum field theory is formulated
 is truncated to a finite number of degrees of
freedom, relativity is lost.  The goal of this paper is to construct 
 mass operators acting on a truncated Fock space for an n-body system.  
Up to a given
truncation the mass operator should determine all the possible reactions
that were allowed by the field theory.  Further the many-body theory
should satisfy the following properties:

-For a given truncation of Fock space the dynamics should arise from a 
many-body mass operator.

-Generators built out of the mass operator (and other kinematic operators)
should satisfy the Poincare commutation relations.

-Forces come from vertex interactions, which generate the mass operator,
 yet should look like relativistic "potentials".

-The theory should be Lorentz covariant.

-Bound state problems are solved as eigenvalue-eigenvector problems, from
which renormalized masses can be
calculated.

The theory to be formulated here is intermediate between a relativistic quantum 
mechanics for a fixed number of particles, and quantum field theory, as a theory
with infinite numbers of degrees of freedom, and incorporating full particle 
production and vacuum structure.  It is reminiscent of earlier models such as
the Lee model \cite{e}, with its mass renormalization and limited particle production.
Though the original Lee model was not relativistically covariant, Fuda \cite{f} has
shown, using the front form, how to make it properly relativistic.  The theory
developed here differs from this generalized Lee model in that it allows for an
arbitrary number of produced particles, up to a prespecified limit, given by
the original truncation.  As will be discussed in the conclusion there are
still open questions about the cluster properties for such a theory.

The procedure for obtaining a mass operator 
 will be carried out in the context of point form relativistic
quantum mechanics \cite{g}, in which all of the interactions are put in the 
four-momentum operator and the Lorentz generators are kinematic.  The
four-momentum operator is written as the product of a mass operator
times a four-velocity operator, which is the Bakamjian-Thomas construction
 in the point form. The interacting part of the mass operator is obtained from
matrix elements of a vertex, products of local field operators evaluated
 at the space-time point zero,
using velocity states, eigenstates of the four-velocity operator, in which
the matrix elements are diagonal in the four-velocity.

In section 2 the elements of point form quantum mechanics are reviewed
and both free and interacting mass operators are constructed, using velocity
states, diagonal in the
overall four-velocity of an n particle state.  In section 3 the techniques 
discussed in
section 2 are applied to the simple example of a scalar "nucleon", scalar
"pion" vertex.
\section{Point Form Relativistic Quantum Mechanics and Mass Operators}
In point form relativistic quantum mechanics all interactions are put in
the four-momentum operator.  The Lorentz generators contain no
interactions, so the point form is manifestly Lorentz covariant.  In fact
the Lorentz generators will almost never be used;  instead global Lorentz
transformations will define the transformation properties of operators
and states.  If $U_\Lambda$ is a unitary operator representing the
Lorentz transformation $\Lambda$, then the four-momentum operator must
satisfy the following point form equations:
\begin{eqnarray}
[P_\mu,P_\nu]&=&0\\
U_\Lambda P_\mu U_\Lambda^{-1}&=&(\Lambda^{-1})_\mu^\nu P_\nu.\
\end{eqnarray}
These equations are simply one way of writing the Poincar\'{e} commutation
relations in which the relations of the four-momentum operators are
emphasized.  The mass operator is defined to be $M:=\sqrt{P\cdot P}$ and
must have a nonnegative spectrum.

Since the four-momentum operators are the generators of space-time
translations, they can be used to define the relativistic generalization
of the time dependent Schr\"{o}dinger
equation,
\begin{eqnarray}
i\frac{\partial \Psi_x}{\partial x^\mu}&=&P_\mu\Psi_x,
\end{eqnarray}
where $\Psi_x$ is an element of the Hilbert space. $x(=x_\mu)$ is a
space-time point and acts as four parameters, playing the same role as the
time parameter in the nonrelativistic time dependent Schr\"{o}dinger equation.
  From Eq.3 it follows that $\Psi_x$ satisfies a
generalized Klein-Gordan equation,
\begin{eqnarray}
(\frac{\partial}{\partial x^\mu}\frac{ \partial}{\partial x_\mu} +M^2)\Psi_x&=&0,
\end{eqnarray}
where $M$ is the mass operator.  It is Eqs.3,4 that take the place of
equations of motion for operators.  If $P_\mu$ has no explicit space-time
dependence then Eq.3 can be written as an eigenvalue equation for the
four-momentum operator:
\begin{eqnarray}
P_\mu\Psi&=&p_\mu\Psi.
\end{eqnarray}

The simplest way in which the point form equations can be satisfied is
with single particle irreducible representations of the Poincar\'{e} group. 
Then the Hilbert space is $L^2(R^3)\bigotimes V^j$, where $V^j$ is the 2$j$+1
dimensional spin space for a particle of spin $j$, and the state $|p,
\sigma>$ transforms under Poincar\'{e} transformations as
\begin{eqnarray}
P_\mu|p,\sigma>&=&p_\mu|p,\sigma>,\\
U_\Lambda|p,\sigma>&=&\sum|\Lambda
p,\sigma^{'}>D^j_{\sigma^{'},\sigma}(R_W),\\
R_W:&=&B^{-1}(\Lambda v)\Lambda B(v).\
\end{eqnarray}
Here the four-momentum $p$ satisfies the mass constraint $p\cdot p=m^2$
and the four-velocity $v$ is defined by $v=p/m$.  $B(v)$ is a boost (see
reference \cite{h}), a Lorentz transformation that takes the rest 
four-momentum,
$p^{rest}=(m,0,0,0)$ to the four-momentum $p$:  $B(v)p^{rest}=p$.  $R_W$
is a Wigner rotation and $D^j( )$ is a matrix element of the rotation
group.  It is clear that the operators Eqs. 6,7 satisfy, by construction,
the point form equations.

Many-particle states with the same transformation properties as the
single particle ones are conveniently obtained by introducing creation
and annihilation operators.  Let $a^\dagger(p,\sigma)$ be the operator
that creates the state $|p,\sigma>$ from the vacuum.  If $a(p,\sigma)$ is
its adjoint, these operators must satisfy the following relations:
\begin{eqnarray}
[a(p,\sigma),a^\dagger(p^{'},\sigma^{'})]_\pm&=&E\delta^3(p-p^{'})\ 
\delta_{\sigma,\sigma^{'}}\\
U_a a^\dagger(p,\sigma)U_a^{-1}&=&e^{ip\cdot a} a^\dagger(p,\sigma),\\
P_\mu(fr)&=&\sum \int\frac{d^3 p}{E} p_\mu a^\dagger(p,\sigma)a(p,\sigma),\\
U_\Lambda a^\dagger(p,\sigma) U_\Lambda^{-1}&=&\sum
a^\dagger(\Lambda p,\sigma^{'}) D^j_{\sigma^{'},\sigma}(R_W).\
\end{eqnarray}
Here $P_\mu(fr)$ is the free four-momentum operator and plays a role
analogous to the free Hamiltonian in nonrelativistic quantum
mechanics.  Again it is straightforward to show that $P_\mu$
satisfies the point form equations.  $U_a$ in Eq.10 is the unitary
operator representing the four-translation $a$.

To prepare for the construction of interacting mass operators, it is
convenient to introduce velocity states, states with simple Lorentz
transformation properties.  If a Lorentz transformation is applied to a
many-particle state,
$|p_1,\sigma_1...p_n,\sigma_n>=a^\dagger(p_1,\sigma_1)
...a^\dagger(p_n,\sigma_n)|0>$, then it is not possible to couple all the
momenta and spins together to form spin or orbital angular momentum
states, because the Wigner rotations for each momentum state are
different.  However, velocity states, defined as $n$-particle states in
their overall rest frame boosted to a four-velocity $v$ will have
the desired Lorentz transformation properties:
\begin{eqnarray}
|v,\vec{k_i},\mu_i>:&=&U_{B(v)}|k_1,\mu_1...k_n,\mu_n>\\
&=&\sum |p_1,\sigma_1...p_n,\sigma_n>\prod
D^{j_i}_{\sigma_i,\mu_i}(R_{W_i}).\\
U_\Lambda|v,\vec{k_i},\mu_i>&=&U_\Lambda U_{B(v)}|k_1,\mu_1...k_n,\mu_n>
\nonumber\\
&=&U_{B(\Lambda v)}U_{R_W}|k_1,\mu_1...k_n,\mu_n>\nonumber\\
&=&\sum|\Lambda v,R_W \vec{k_i},\mu_i^{'}>\prod
D^{j_i}_{\mu_i^{'},\mu_i}(R_W).\
\end{eqnarray}
Unlike the Lorentz transformation of an $n$-particle state, where all the
Wigner rotations of the D functions are different, in Eq.15 it is seen
that the Wigner rotations in the D functions are all the same and given
by Eq.8.  Moreover the same Wigner rotation also multiplies the internal
momentum vectors, which means that for velocity states, spin and orbital
angular momentum can be coupled together exactly as is done
nonrelativistically (for more details see reference \cite{h}).  The relationship
between single particle and internal momenta is given by
$p_i=B(v)k_i,\sum\vec{k_i}=0$; Eq.14 provides the link between velocity and
 many-particle states, where $R_{W_i}$  is given by
replacing
$p$ by $k_i$ and $\Lambda$ by
$B(v)$ in Eq.8.  From the definition of velocity states it then follows
that
\begin{eqnarray}
V_\mu|v,\vec{k_i},\mu_i>&=&v_\mu|v,\vec{k_i},\mu_i>,\\
M(fr)|v,\vec{k_i},\mu_i>&=&m_n
|v,\vec{k_i},\mu_i>,\\
P_\mu(fr)|v,\vec{k_i},\mu_i>&=&m_nv^\mu|v,\vec{k_i},\mu_i>,\
\end{eqnarray}
with  $m_n=\sum\sqrt{m_i^2 +\vec{k_i}^2}$ the 'mass' of the $n$-particle
velocity state and $P_\mu(fr)=M(fr)V_\mu$.  On velocity states the free
four-momentum operator has been written as the product of the 
four-velocity operator times the free mass operator, which is the
Bakamjian-Thomas construction for the point form.  More generally four-momentum
 operators are written as $P_\mu=M V_\mu$,
where the four-velocity operator is defined by $V_\mu:=\frac{P_\mu(fr)}{M_{fr}}$;
$V_\mu$ is diagonal on velocity states, as seen in Eq.16.  The mass operator is
the sum of free and interacting mass operators, $M=M_{fr}+M_I$;  if the mass 
operator commutes with  the four-velocity operator and Lorentz transformations,
then the point form Eqs.1,2 will be satisfied.

The goal now is to construct an interacting mass operator on a truncated Fock
 space, where the Fock space is formed from creation and annihilation operators.
The truncated Fock space is the space of the many-body system, a system with a 
finite number of degrees of freedom.  For example if the creation and annihilation
operators are those for (bare) pions and nucleons, the truncated Fock space 
could be chosen as the direct sum of two nucleon spaces with zero to n pions, 
appropriate for the scattering of two nucleons as well as the production of 
pi mesons.

The same creation and annihilation operators can also be used to form local
fields;  the general procedure for this construction is given in reference 
\cite{i} and is reviewed in detail in one of the papers in this series, 
dealing with constructing current operators for arbitrary spin particles
\cite{j}.  Moreover reference \cite{i} shows
how to take products of such arbitrary spin fields and couple them together
 to form Lorentz scalar densities, which, following this reference  will be 
denoted by $\mathcal{H}(x)$.  Examples of such scalar densities of products of
local fields are the familiar $\bar{\Psi}\gamma _5 \Psi\phi$ for pion-nucleon
vertices and $\bar{\Psi}\gamma_{\mu}\Psi A^{\mu}$ for electron-photon vertices.

The fact that $\mathcal{H}(x)$ is a Lorentz scalar density means that
$U_\Lambda \mathcal{H}(x)U_{\Lambda}^{-1}=\mathcal{H}(\Lambda x)$.  Therefore
at the space-time point $x=0$, $U_\Lambda\mathcal{H}(0)U_\Lambda ^{-1}=
\mathcal{H}(0)$, so that $\mathcal{H}(0)$ is a Lorentz scalar.  The velocity
 state matrix element of $\mathcal{H}(0)$ on the truncated Fock space,
 $<v^{'}\vec{k}_i^{'}\mu_{i}^{'}|\mathcal{H}(0)|v\vec{k}_i\mu_i>|_{v^{'}=v}$,
where the initial and final four-velocities are the same,
can be viewed as the kernel of an interacting mass operator on the truncated
Fock space, because it commutes with the four-velocity operator (since it is
diagonal in the four-velocity, by construction) and commutes with Lorentz 
transformations.

However, since $\mathcal{H}(0)$ is constructed from products of fields, its 
velocity state matrix elements may not give kernels of well defined operators
on the truncated Fock space.  In order that the interacting mass operator be
well defined on the truncated Fock space, matrix elements of $\mathcal{H}(0)$ are
multiplied by "form factors", functions of $(p^{'}-p)^2=(m^{'}v^{'}-mv)^2$,
where $m$ and $m^{'}$ are the free masses of the many-particle state,
given in Eq.17.  But since the initial and final four-velocities are the same,
the form factor will be a function of the magnitude of mass differences only, 
namely $\Delta m:=|m^{'}-m|$.  The definition of the interacting many-body mass
operator is then given by
\begin{eqnarray}
M_I:&=&<v^{'},\vec{k}^{'}_i,\mu_i^{'}|\mathcal{H}(0)|
v,\vec{k}_i,\mu_i>|_{v^{'}=v}f(\Delta m),
\end{eqnarray}
and by virtue of the form factor $f(\Delta m)$ is a well defined operator on
the truncated Fock space.  It commutes with Lorentz transformations and the
four-velocity operator, so that $P_\mu=(M_{fr}+M_I)V_{\mu}$ satisfies the 
point form equations, Eq.1,2 and provides the starting point for a relativistic
many-body theory.

It should be noted that $M_I$ actually is independent of the four-velocity:
\begin{eqnarray}
M_I&=&<v^{'},\vec{k}^{'}_i,\mu^{'}_i|\mathcal{H}(0)|v,\vec{k}_i,\mu_i>|_{v^{'}=v}
f(\Delta m)\nonumber\\
&=&<k_1^{'}\mu_1^{'}...|U_{B(v)}^{-1}\mathcal{H}(0)U_{B(v)}|k_1\mu_1...>|_{v^{'}=v}
f(\Delta m)\nonumber\\
&=&<k_1^{'}\mu_1^{'}...|\mathcal{H}(0)|k_1\mu_1...>|_{v^{'}=v}f(\Delta m),
\end{eqnarray}
where the $k_i$ and $k_i^{'}$ are the internal momenta defined in Eq.13 and
use has been made of the fact that $\mathcal{H}(0)$ is a Lorentz scalar.  Also,
since under Lorentz transformations the internal momenta are rotated by the 
Wigner rotation of Eq.15 the mass operator is rotationally invariant and satisfies
all the properties of a "potential".  Finally, since $M_I$ arises from products
of fields, it always has only off-diagonal matrix elements in the truncated
Hilbert space.

\section{An Example:  The Scalar Nucleon-Pion Vertex}
To show how mass operators and generalized Lippmann-Schwinger equations
result from vertices, this section works out the example
of a scalar "nucleon" interacting with a scalar "pion".  The vertex is of
 the form $\mathcal{H}(x)=g\Psi^\dagger(x)\Psi(x)\phi(x)$ where
$g$ is a coupling constant.  The interacting mass operator can be written from
Eq.19 as
\begin{eqnarray}
M_I&=&gf(\Delta
m)<v,\vec{k_i},\mu_i|\Psi^\dagger(0)\Psi(0)\phi(0)|v,\vec{k_i^{'}},
\mu_i^{'}>\nonumber,\
\end{eqnarray}
and various baryon number sectors evaluated.

As a first truncation  consider the baryon number one sector, with the
Hilbert space the direct sum of bare nucleon plus bare nucleon and bare
pion.  Then the total mass operator can be written as a two by two matrix
operator in the direct sum space:
\begin{eqnarray}
M&=&\left[\begin{array}{cc}m_N(0)&gK^\dagger\\gK&D_2(0)
\end{array}\right]\
\end{eqnarray}
where the internal momentum variables are defined by\\
$p_N=B(v)(\omega_N(0),\vec{k})$ and
$p_\pi=B(v)(\omega_\pi,-\vec{ k})$ with
$\omega_N(0)=\sqrt{m_N^2(0)+\vec{k}^2}$
and $\omega_\pi=\sqrt{m_\pi^2+\vec{k}^2}$. 
$m_N(0)$ is the bare nucleon mass and\\ 
$D_2(0):=\sqrt{m_N^2(0)+\vec{k}^2}
+\sqrt{m_\pi^2+\vec{k}^2}=\omega_N(0)+\omega_\pi$, a diagonal operator in
the two particle space, is the relativistic two particle mass. The
interacting mass operator that connects the two particle space to the one
nucleon space is the kernel
\begin{eqnarray}
<\vec{k}|K|>&=&f(\Delta m)\nonumber\\&=&f(\omega_N(0)+\omega_\pi-m_N(0)).\
\end{eqnarray}

On the direct sum space the problem to be solved is the mass operator
eigenvalue-eigenvector problem, $M\phi=m\phi$, where, for the bound state
problem, $m=m_N$ is the physical nucleon mass, while for the scattering
problem, $m=\sqrt{s}$ is the invariant relativistic energy for pion-nucleon
scattering. Designating the components of the physical one nucleon state
as
$(\phi_1^N,
\phi_2^N)$, the mass operator eigenvalue equation becomes
\begin{eqnarray}
m_N(0)\phi_1^N+g(K,\phi_2^N)&=&m_N\phi_1^N\\
gK\phi_1^N+D_2(0)\phi_2^N&=&m_N\phi_2^N\\
\phi_2^N&=&g(m_N-D_2(0))^{-1}K\phi_1^N\\
g^2(K,(m_N-D_2(0))^{-1}K)&=&m_N-m_N(0),\
\end{eqnarray}
where in the last equation, $\phi_1^N$ has been cancelled  from both
sides of Eq.26 because it is a number (to be determined by normalization
requirements).  For a given truncation, Eq.26 provides the relationship
between the bare and physical nucleon mass.  If $m_N$ is taken to be the
physical nucleon mass, then, for a given form factor, the bare nucleon mass,
 $m_N(0)$, is fixed. 
Written out more explictly, Eq.26 becomes
\begin{eqnarray}
g^2\int \frac{d^3 k}{\omega_N(0)\omega_\pi}
\frac{|f(\omega_N(0)+\omega_\pi-m_N(0))|^2}{m_N-\omega_N(0)-\omega_\pi}&=&
m_N-m_N(0),\
\end{eqnarray}
and $f$ is used to regularize the large momentum components in the
integral.  The nucleon wave function is normalized to one; that is, 
$|\phi_1^N|^2+(\phi_2^N,\phi_2^N)=1$, where $\phi_2^N$ is given in Eq.25.

To formulate the pion-nucleon scattering problem, the mass operator, Eq.21,
is written as the sum of the free mass operator, $M_0$, and the
interacting mass operator, $M_I$, $M=M_0 +M_I$, where
\begin{eqnarray}
M_0&=&\left[\begin{array}{cc}m_N(0)&0\\0&D_2(0)
\end{array}\right]\,\\
M_I&=&\left[\begin{array}{cc}0&gK^{\dagger}\\gK&0
\end{array}\right]\,\
\end{eqnarray}
The relativistic Lippmann-Schwinger equation can then be written as an
integral equation on the direct sum space with scattering wave function
$\Psi=(\Psi_1,\Psi_2)$:
\begin{eqnarray}
\Psi&=&\Phi+(\sqrt{s}-M_0+i\epsilon)^{-1}M_I\Psi\\
\Psi_1&=&\frac{g}{\sqrt{s}-m_N(0)}(K,\Psi_2)\\
\Psi_2&=&\phi_2+
\frac{g^2}{\sqrt{s}-m_N(0)}(\sqrt{s}-D_2(0))^{-1}K(K,\Psi_2),
\end{eqnarray}
where the second term in Eq.32 looks like a separable potential with a
coupling constant $g^2/(\sqrt{s}-m_N(0))$.  That is, in this simple
truncation the Lippmann-Schwinger equation can be solved exactly to
compute the pion-nucleon scattering amplitude.  $\Phi=(0,\phi_2)$, where 
$\phi_2$ is the free pion-nucleon momentum state, and all $\sqrt{s}$
factors in Eqs.30,31 should be replaced by $\sqrt{s}+i\epsilon$.

The function $f(\Delta m)$ used to relate bare and physical mass, Eq.27
and generate the separable potential, Eq.32, can also be used in other
baryon number sectors.  For baryon number two, consider the truncated
space consisting of two bare nucleons plus two bare nucleons and bare
pion, a direct sum of two particle and three particle spaces.  In this
case the mass operator has the form
\begin{eqnarray}
M&=&\left[\begin{array}{cc}D_2(0)&gK^{\dagger}\\gK&D_3(0)
\end{array}\right]\,
\end{eqnarray}
where the internal momenta are related to single particle momenta by the
following relations: $p_1^{'}=B(v)(\omega,\vec{k}),
p_2^{'}=B(v)(\omega,-\vec{k})$, with $\omega=\sqrt{m_N^2(0)+\vec{k}^2}$;
$p_1=B(v)(\omega_1,\vec{k}_1), p_2=B(v)(\omega_2,\vec{k}_2), p_\pi=B(v)
(\omega_{\pi},\vec{k}_\pi)$, with $\omega_i=\sqrt{m_N^2(0)+\vec{k}_i^2}$
and $\vec{k}_1+\vec{k}_2+\vec{k}_\pi=0$.  The diagonal operator
$D_2(0)=2\omega$ is the bare two particle relativistic mass, while the
diagonal operator $D_3(0)=\omega_1+\omega_2+\omega_\pi$ is the bare three
particle mass.  The two particle to three particle transition kernel is
given by
\begin{eqnarray}
<\vec{k}_1\vec{k}_2|K|\vec{k}>&=&f(\omega_1+\omega_2+\omega_\pi-2\omega)
\nonumber\\
&&(\delta^3(\vec{k}+\vec{k}_1)+\delta^3(\vec{k}-\vec{k}_1)+\delta(\vec{k}+
\vec{k}_2)+\delta(\vec{k}-\vec{k}_2)),\
\end{eqnarray}
where the argument in $f$ is (the magnitude of) the mass difference
between the three and two particle states.  The four delta functions are
a consequence of the two nucleons being identical particles.

With the mass operator given in Eq.33 a "deuteron" bound state can be
calculated as a two component mass operator eigenvector, corresponding to
the deuteron mass $m_D$:
\begin{eqnarray}
M\phi^D&=&m_D\phi^D,\\
D_2(0)\phi_2^D+gK^{\dagger}\phi^D_3&=&m_D\phi_2^D\\
gK\phi_2^D+D_3(0)\phi_3^D&=&m_D\phi_3^D\\
\phi_3^D&=&g(m_D-D_3(0))^{-1}K\phi_2^D\\
D_2(0)\phi_2^D+g^2K^{\dagger}(m_D-D_3(0))^{-1}K\phi_2^D&=&m_D\phi_2^D.\
\end{eqnarray}
Though the first term in Eq.39 is the bare relativistic kinetic energy,
the second term, which looks like a potential energy term, contains the
unknown deuteron mass eigenvalue; 
because the component of the deuteron wavefunction in the three particle
sector has been eliminated, Eq.39 is not of the usual form of an
eigenvalue equation, but is known in coupled channel problems. 
 The deuteron wave function is normalized by writing
$||\phi_2^D||^2+||\phi_3^D||^2=1$.

The mass
operator, Eq.33 can be used to calculate scattering in the
nucleon-nucleon and nucleon-nucleon-pion channels by
again writing the mass operator as a sum of the free mass operator
plus interacting mass operator,
$M=M_0+M_I$:
\begin{eqnarray}
M_0&=&\left[\begin{array}{cc}D_2(0)&0\\0&D_3(0)
\end{array}\right]\,\\
M_I&=&\left[\begin{array}{cc}
0&gK^{\dagger}\\gK&0
\end{array}\right]\,\\
\Psi&=&\Phi+(\sqrt{s}-M_0+i\epsilon)^{-1}M_I\Psi;\\
\Psi_2&=&\phi_2+(\sqrt{s}-D_2(0))^{-1}
gK^{\dagger}\Psi_3,\\
\Psi_3&=&(\sqrt{s}-D_3(0))^{-1}gK\Psi_2\\
\Psi_2&=&\phi_2+g^2(\sqrt{s}-D_2(0))^{-1} K^{\dagger}
(\sqrt{s}-D_3(0))^{-1}
K\Psi_2,\
\end{eqnarray}
where $\Phi=(\phi_2,0)$, and the driving term is the two free
nucleon momentum state, $\phi_2=\delta^3(\vec{k}-\vec{k}_{initial})$. 
Again Eq.45 has the form of a driving term plus "potential", from which
it is possible to calculate scattering amplitudes for the two processes
mentioned above.

Finally, the baryon zero sector gives the ground state and physical pion
mass.  However in the truncation where there are no nucleon-antinucleon
pairs, and only the direct sum of Fock vacuum, and one or more bare pion
states, there is no pion mass renormalization and the physical vacuum is
the Fock vacuum.  If however a nucleon-antinucleon two particle space is
added to the Fock vacuum along with a nucleon-antinucleon-pion space and a
one and two pion space, the mass operator will have off-diagonal terms
linking the Fock vacuum to the nucleon-antinucleon-pion Hilbert space,
and  pion mass renormalization will occur, as well as pion-pion
scattering.

More generally for a given truncation the mass operator will be a
quintdiagonal matrix operator for which bound and scattering states can
be calculated.  If the function $f$ occurring in the off-diagonal kernal
is fixed, then for the given truncation all channels will be determined
by $f$.
\section{Conclusion}
This paper has shown how to compute interacting mass operators from a
given vertex, by multiplying velocity state matrix elements of the vertex 
operator by a form factor that makes the mass operator a well defined operator
 on the truncated Fock space appropriate for a few-body system.  The total mass
operator, the sum of free and interacting mass operators, has a discrete 
spectrum that gives the bound states, and continuous spectrum the scattering 
states.  And just as the total momentum and spin projection are extracted from
an isolated many-body nonrelativistic system, so in the point form the overall
four-velocity and spin projection are extracted from an isolated many-body
relativistic system.  Since the interacting mass operator comes from a vertex
operator, the interactions are all due to the exchange of particles.  Yet, as
shown in section 3 the coupled channel problem can be rewritten so that the
relativistic Lippmann-Schwinger equations contain relativistic "potentials".

There are a number of applications of the formalism discussed here.  The most
obvious is a more realistic treatment of the pion-nucleon system, in which the
spins are correctly taken into account.  Then the interacting mass operator
will contain spinors, but, as can be seen from the way in which fields for 
arbitrary spin particles can be coupled together to form Lorentz scalar 
densities \cite{i}, the coupling can also include the spin 3/2 delta resonance.
Lee and coworkers \cite{k}, as well as Fuda \cite{l} have written T matrix 
equations for pion-nucleon scattering, using basically the instant form of
relativistic quantum mechanics, in which vertices are used to generate
interactions in the Hamiltonian.  A similar procedure should be possible in the
point form.

A second application is to the study of the meson and baryon spectrum, arising 
as bound states of quarks.  While the work of Plessas et al \cite{m} can be 
interpreted as a point form mass operator, the hyperfine part of the mass
operator only indirectly comes from a vertex.  Closer in spirit to what is 
developed here is the dissertation of Krassnigg \cite{n} who studies the
vector mesons as bound states of quark-antiquark pairs.  The truncated Hilbert
space is the direct sum of quark-antiquark and quark-antiquark, pseudoscalar
meson spaces, and the hyperfine interaction is given from a vertex of the form
$\mathcal{H}(x)=\bar{\Psi}(x)\gamma_5\Psi(x)\phi(x)$, where $\bar{\Psi}(x)$ is
the quark field and $\phi(x)$ the pseudoscalar meson field.  The main difference
with the mass operator discussed in this work is that the free mass operator is
replaced by a harmonic oscillator mass operator to simulate quark confinement.
But otherwise the coupled channel equations are very similar to those of 
section 3, and result, for example, in excited states having  widths
and shifted masses.

For a given truncation there is a definite mass renormalization, which
however will change if the truncation changes.  Already at this point 
violations of cluster properties will occur, for if a set of bare particles
are moved far away from another set, the mass renormalization for the
separated set will not be the same as with all the particles.  But for a
given truncation and mass renormalization, it should be possible to construct 
unitary operators (packing operators,see reference \cite{d}) that give the
 desired cluster properties
for the many-body system.  Further there is reason to believe that the packing
operators for vertex interactions should be relatively straightforward to
calculate.  The desirable cluster properties come about because
$f$ has as an argument the difference of the relativistic
masses of the initial and final sets of particles.  The mass operator is the
product of this function multiplied by velocity state matrix elements of
the vertex operator.  For a given vertex only three particles
interact while the momentum of the others are unchanged, 
resulting in the cancellation of energies in the argument of $f$. 
 Then the argument of $f$ depends only on the momentum of the
interacting particles and not the others. Because of this peculiar feature of
the argument of $f$ in the point form, the cluster properties may be relatively 
simple.  But this remains to be investigated.

More generally as pointed out in the introduction, the coupled channel mass 
operators that result from vertex interactions allow for a limited amount of
 particle production,
depending on the truncation chosen.  Such theories are intermediate between
relativistic direct interaction theories for a fixed number of particles and
quantum field theories with infinite degrees of freedom.  Moreover, quantum
field theories satisfy cluster separability properties (see for example,
\cite{i}), while relativistic direct interaction theories for a fixed number of 
particles require packing operators to satisfy cluster properties \cite{d}.  
While the relativistic coupled channel theory discussed in this paper has
limited particle production and requires mass renormalization, the way mass
renormalization depends on cluster properties remains an open question.  This
issue is particularly relevant in light of a theorem by Aks \cite{o} that says
in a  four dimensional quantum field theory particle production and vacuum 
polarization necessarily go together.

The procedures given in this paper for constructing
strongly interacting mass operators from vertices can also be applied to 
electromagnetic vertices, where $\mathcal{H}(0)$ is of the form
 $J_\mu(0)A^{\mu}(0)$;  in the following papers current operators are
constructed for particles with arbitrary spin and form factors \cite{j}, as well
 as analyzing the Gupta-Bleuler formalism appropriate for the above coupling 
\cite{p}.  
Finally the last paper in this series \cite{q} shows how the free current operator is
modified from its one-body form to generate the appropriate many-body current
operator in the presence of interactions.  The goal then, in this series of
papers is to see whether the nonrelativistic methods that
have been so successful in low energy nuclear physics, where the Hamiltonian is
the sum of matter, photon, and electromagnetic coupling Hamiltonians,
 can be generalized to higher energies where relativity is required and the
 mass operator is similarly the sum of matter, photon and electromagnetic
coupling mass operators.
\section{Appendix}
In this appendix the interacting mass operator is obtained from $\mathcal{H}(x)$
by integrating over the forward hyperboloid to get the interacting four-momentum
operator. Let $\mathcal{H}(x)$
be a polynomial in free fields which is a Lorentz scalar density,
$U_\Lambda \mathcal{H}(x) U_\Lambda^{-1}=\mathcal{H}(\Lambda x)$. Then
an interacting four-momentum operator which satisfies the point form
equations can be defined by integrating
$\mathcal{H}(x)$ over the forward hyperboloid:
\begin{eqnarray}
P_\mu(I):&=&\int d^4 x\delta(x\cdot x-\tau^2
)x_\mu\theta(x_0)\mathcal{H}(x);\\ {[P_\mu(I),P_\nu(I)]}&=&0.\
\end{eqnarray}
The components of the interacting four-momentum operator commute with one
another since the commutator of $\mathcal{H}(x)$ with $\mathcal{H}(y)$
is zero if the space-time points $x$ and $y$ are space-like separated.  But
both these points lie on the same forward hyperboloid specified by $\tau$
and hence are space-like separated.  It also follows from Eq.46 that
$P_\mu(I)$ transforms as a Lorentz four-vector, since $\mathcal{H}(x)$
transforms as a Lorentz scalar density.  Hence the interacting
 four-momentum operator satisfies the point form equations, Eq.1,2.

The total hadronic four-momentum operator is the sum of free and interacting
four-momentum operators, $P_\mu(h)=P_\mu(fr)+P_\mu(I)$.  Again it can be
shown that the total four-momentum operator satisfies the point form
equations.  The Lorentz transformation part follows from the fact that
both terms transform as four-vectors.  To show that the components of the
total four-momentum operator commute, it suffices to note that the
components of the free part commute among themselves (from Eq.11), as do
the components of the interacting part (Eq.47).  Thus what must be shown
is that the sum of the cross term commutators give zero.  Now from the definition of
$\mathcal{H}(x)$ as a polynomial in free fields, it follows that $U_a
\mathcal{H}(x)U_a^{-1}=\mathcal{H}(x+a)$.  But $U_a$ is the
exponential of
$P(fr)\cdot a$, so if
$a$ is made infinitesimal, then
\begin{eqnarray}
{[P_\nu(fr),P_\mu(I)]}&=&\int d^4 x\delta(x\cdot x -\tau^2)\theta(x_0) 
x_\mu\frac{\partial}{\partial x^\nu} \mathcal{H}(x);\\
{[P_\mu(fr),P_\nu(I)]}&&\nonumber\\
+[P_\mu(I),P_\nu(fr)]&=&[P_\mu(fr),P_\nu(I)]-[P_\nu(fr),
P_\mu(I)]\nonumber\\
&=&\int d^4 x \delta(x\cdot x-\tau^2)\theta(x_0)(x_\nu\frac{\partial}{\partial
x^\mu} -x_\mu\frac{\partial}{
\partial x^\nu})\mathcal{H}(x)\nonumber\\
&=&0,\
\end{eqnarray}
since $\mathcal{H}(x)$ is a Lorentz scalar density.  Thus the total
four-momentum operator satisfies the point form equations.

Given the total four-momentum operator, the sum of Eqs. 11,46,
 bound and scattering states could in
principle be found by solving the eigenvalue equation, Eq.5.  But the interacting
four-momentum operator defined in Eq.46 is not a well defined operator because
it involves the product of local field operators at the same space-time point.
Moreover, if a restriction to a finite degree of freedom system is made,
the theory is no longer properly relativistic because Eq.47 is not valid on a
truncated space.  Finally, the interacting four-momentum operator given in
Eq.47 is never of Bakamjian-Thomas form;  that is, it can never be written as
the product of a mass operator times a four-velocity operator.
Nevertheless, a Bakamjian-Thomas interacting mass operator can be constructed
from Eq.46 by examining matrix elements of the
interacting four-momentum operator, where the states are velocity states with
the same initial and final four-velocities:
\begin{eqnarray}
<v,\vec{k_i},\mu_i|P_\mu(I)|v,\vec{k^{'}_i},\mu_i^{'}>&=&<v,|\vec{k_i},\mu_i|
\mathcal{H}(0)|v,\vec{k_i^{'}},\mu^{'}_i>\nonumber\\
&&\int d^4 x\delta(x\cdot x-\tau^2)x_\mu\theta(x_0)e^{i(\Delta m)v\cdot x},
\end{eqnarray}
where $\Delta m$ is $m-m{'}$ and $m$ (respectively $m^{'}$) is the mass of the
velocity state, as given in Eq.17. $(\Delta m)v$ is always a timelike four-vector,
although the sign of the time component may be positive or negative, depending 
on the sign of $\Delta m$.

The integral in Eq.50 is a special function and is evaluated in the
following paragraph. The relevant point is that, if the initial and
 final four-velocities 
are arbitrary, the integral is not diagonal in the four-velocity, which is 
another way of saying that an interacting four-momentum operator in quantum
field theory is never of Bakamjian-Thomas form.

But Eq.50 can be used to construct an interacting mass 
operator, generated from vertex interactions, by defining
\begin{eqnarray}
M_I&=&<v\vec{k}_i\mu_i|\mathcal{H}(0)|v\vec{k}_i^{'}\mu_i^{'}>f(\Delta m),
\end{eqnarray}
where $f$ is the function defined in Eq.50 and evaluated in the following 
paragraph.  If it is replaced with an arbitrary form factor, the mass operator
is the same as that given in Eq.50, which was the starting point for a many-body
theory.

 The integral in Eq.50 can be computed as follows.  Define the Lorentz invariant
function $I^{\pm}_{\tau}$ as
\begin{eqnarray}
I^{\pm}_{\tau}(p)&=&\int d^4 x\delta(x\cdot x-\tau^2)\theta(x_0)
e^{{\pm}ip\cdot x}\\
I^\pm_{\tau}(\Lambda p)&=&I^\pm_{\tau}(p),\\
I^\pm_{\tau}(sp)&=&s^{-2}I^\pm_{s\tau}(p)
\end{eqnarray}
 From reference \cite{r}
the function $I^{\pm}_\tau(p)$ is given by
\begin{eqnarray}
I^+_\tau(p)&=&\pi^2\tau(p\cdot p)^{-1/2}[N_1(\tau\sqrt{p\cdot p})-i
\epsilon(p_0)J_1(\tau\sqrt{p\cdot p})]\\
I_\tau^{-}(p)&=&(I^{+}_\tau(p))^{\ast},\
\end{eqnarray}
for $p$ timelike.

Then for $\Delta m>0$,
\begin{eqnarray}
\int d^4 x\delta(x\cdot x-\tau^2)x_\mu \theta(x_0)e^{i(\Delta m)v\cdot x}
&=&(i\Delta m )^{-1}\partial/\partial v^\mu I^+_\tau(\Delta mv)\nonumber \\
&=&-i(\Delta m)^{-3} \partial/\partial v^\mu 
I^+_{\Delta m\tau}(v)\nonumber\\
&=&v^\mu f_\tau(\Delta m) ,\ 
\end{eqnarray}
where $f_\tau(\Delta m)=-\pi^2\tau(\Delta m)^{-2}\alpha^{-1}\partial/\partial\alpha
[\alpha^{-1}J_1(\Delta m\tau\alpha)+i\alpha^{-1}N_1(\Delta m\tau\alpha)]
_{\alpha=1}$ with $\alpha:=\sqrt{v\cdot v}$.  For $\Delta m<0$, the desired 
integral is
\begin{eqnarray}
\int d^4 x\delta(x\cdot x-\tau^2)x_\mu\theta(x_0)e^{-i|\Delta m|v\cdot x}
&=&i|\Delta m|^{-3}\partial/\partial v^\mu I^-_{|\Delta m|\tau}(v)\nonumber\\
&=&v^\mu f^{\ast}_{\tau}(|\Delta m|).
\end{eqnarray}
The integrals in Eqs.57,58 are complex conjugates of one another, with both
depending on the magnitude of the difference between final and initial
 masses.  $J_1$ and $N_1$ are various forms of Bessel functions, as given in
reference \cite{r}.

\end{document}